\documentclass[12pt]{article}
\usepackage{graphicx}
\usepackage{amsmath} 
\usepackage{amsfonts} 
\usepackage{amssymb} 
\usepackage{amsthm} 
\usepackage{color} 

\def\London{Particle Physics Research Center\\
Queen Mary University, London, UK}

\def\Title#1{\begin{center} {\Large #1 } \end{center}}
\def\Author#1{\begin{center}{ \sc #1} \end{center}}
\def\Address#1{\begin{center}{ \it #1} \end{center}}

\newenvironment{Abstract}{\begin{quotation}  }{\end{quotation}}
\newenvironment{Presented}{\begin{quotation} \begin{center} 
             PRESENTED AT\end{center}\bigskip 
      \begin{center}\begin{large}}{\end{large}\end{center} \end{quotation}}

\begin{document}
\begin{titlepage}

\vfill
\Title{Hyper-Kamiokande\\Towards a measurement of CP violation in lepton sector}
\vfill
\Author{Stephanie Bron}
\Address{\London}
\vfill
\begin{Abstract}
We present the latest Hyper-Kamiokande sensitivity study showing that, with a total exposure of 13\,MW$\times 10^{7}$ seconds integrated beam power, the CP phase,
$\delta_{CP}$, can be determined better than 21 degrees for all possible values of $\delta_{CP}$ and that CP violation can be established with a significance of more
than 3$\sigma$ (5$\sigma$) for 78\% (62\%) of the $\delta_{CP}$ parameter space. 
\end{Abstract}
\vfill
\begin{Presented}
NuPhys2015, Prospects in Neutrino Physics\\
Barbican Centre, London, UK,  December 16--18, 2015
\end{Presented}
\vfill
\end{titlepage}
\def\thefootnote{\fnsymbol{footnote}}
\setcounter{footnote}{0}

\section{Introduction}

CP-symmetry states that the laws of physics are invariant under the combination of charge conjugation and parity transformations. Thus, if CP holds, the oscillation 
probabilities for $\nu$ and $\bar{\nu}$ should be the same.
The oscillation probability depends on the PMNS matrix:

\begin{equation}
\label{PMNS}
 U =  \begin{pmatrix}
	      1 & 0 	 & 0	\\
	      0 & c_{23} & s_{23}\\
	      0 & -s_{23}& c_{23}\\
	      
      \end{pmatrix}
      \begin{pmatrix}
	      c_{13} 			& 0 		& s_{13}e^{-i\delta_{CP}}	\\
	      0 			& 1		& 0				\\
	      s_{13}e^{-i\delta_{CP}} 	& 0 		& c_{13}			\\
	      
      \end{pmatrix}
      \begin{pmatrix}
	      c_{12} 	& s_{12} & 0	\\
	      -s_{12} 	& c_{12} & 0	\\
	      0 	& 0	 & 1	\\
	      
      \end{pmatrix},
\end{equation}

where $c_{ij}\equiv \cos(\theta_{ij})$, $s_{ij}\equiv \sin(\theta_{ij})$ and $\delta_{CP}$ is the Dirac type CP phase in the lepton sector.

If the complex phase, $\delta_{CP}$, in the PMNS matrix is non-zero, CP-symmetry is violated. 

\begin{figure}[h!]
\begin{center}
\includegraphics[scale=0.5]{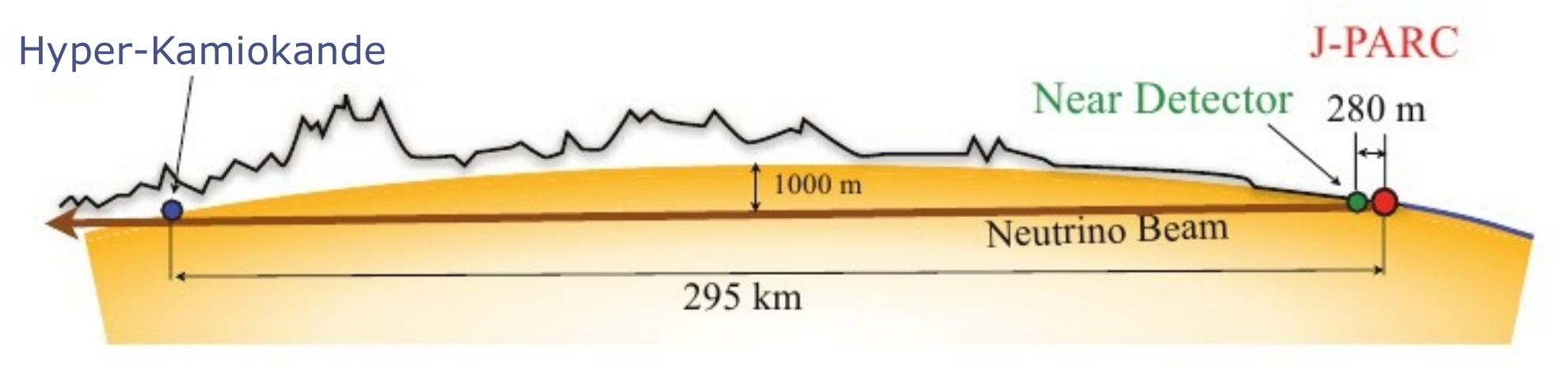}
\end{center}
\caption{Overview of the long baseline experiment, Hyper-Kamiokande.}
\label{layout}
\end{figure}

To test this assumption, we need an experiment sensitive to neutrino oscillations including an accelerator producing beams mainly composed of either $\nu$ or 
$\bar{\nu}$, large statistics and low systematic uncertainties. Our best candidate is Hyper-Kamiokande~\cite{group2014long}, 
a next generation water Cherenkov neutrino detector successor of Super-Kamiokande, that will be used as a far detector for the neutrino beam produced in J-PARC, 295\,km East from Hyper-Kamiokande; this configuration is shown in 
Figure \ref{layout}.

\begin{figure}[h!]
\begin{center}
\includegraphics[scale=0.25]{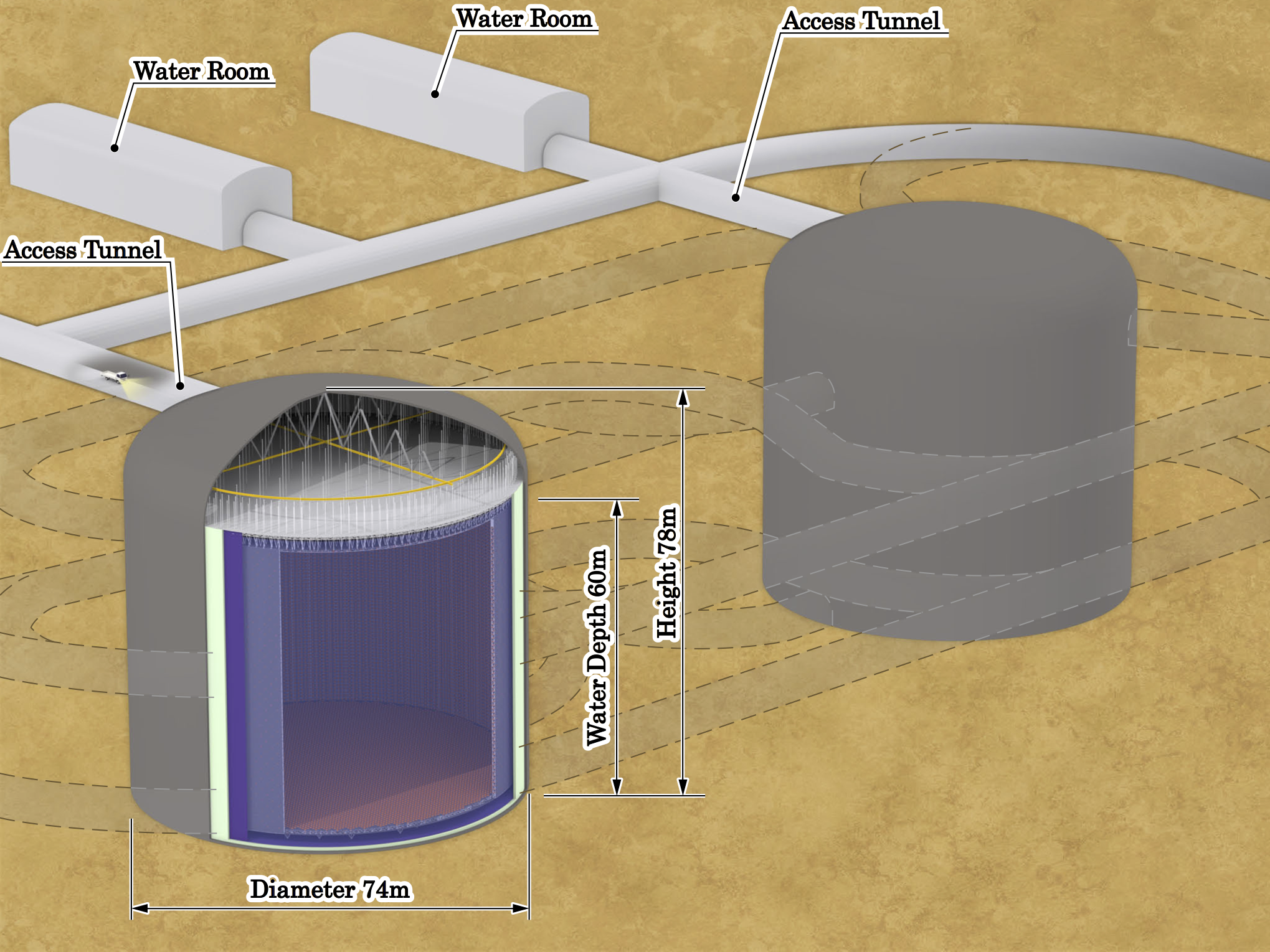}
\end{center}
\caption{Hyper-Kamiokande's design with two tanks.}
\label{design}
\end{figure}

The new Hyper-Kamiokande's design, shown in Figure \ref{design}, consists of two tanks deployed in two stages, i.e. the experiment will start with only one tank and after six years the second tank will be added. The sensitivity studies are presented for a total run time of 10\,years. Each tank has a height of 60\,m and a diameter of 74\,m. The total fiducial volume for the two tanks is 374\,ktons. 

Hyper-Kamiokande's sensitivity to CP violation has been studied using a framework, the Simple Fitter, first developed for the sensitivity study by T2K presented in Ref.~\cite{Abe2015}. 

\section{Analysis Method}

As an appearance measurement is particularly sensitive to $\sin^{2}2\theta_{13}$ and $\delta_{CP}$ while a disappearance measurement is particularly sensitive to $\
\sin^{2}2\theta_{23}$ and $\Delta m^{2}_{23}$, analysing $E_{\nu_{e}}^{rec}$ or $E_{\nu_{\mu}}^{rec}$ gives constraints on the first or the second set of parameters, 
respectively. However, if the two measurements are combined, additional constraints can be obtained and thus a better performance is reached in determining the unknown 
parameters. This is exactly what the Simple Fitter does.

Combined three-flavour appearance, disappearance, $\nu$ mode and $\bar{\nu}$ mode fits are performed and uncertainties on all four parameters are taken into account. 
For these fits we assume that $\sin^{2}2\theta_{13}$, $\delta_{CP}$, $\sin^{2}2\theta_{23}$ and $\Delta m^{2}_{23}$ are unknown. Moreover, the mass hierarchy is taken into 
account as true normal hierarchy (NH) or inverted hierarchy (IH). It is also possible to set the mass hierarchy to be unknown and observe the 
effect on the sensitivity; this strongly reduces the precision, unless further information coming from analysing the atmospheric neutrinos is taken into account. However, as Hyper-K will begin data taking in about 10 years, it is reasonable to assume that the mass hierarchy 
will have been determined by other experiments. Oscillation parameters are treated as shown in Table 1.

\begin{table}[h]
\begin{center}
\begin{tabular}{ccc}
\hline \hline
 Parameter & Nominal value  &  Treatment \\
 \hline
 $\sin^{2}2\theta_{13}$ & 0.10 &  Fitted \\
 $\delta_{CP}$ & 0 &  Fitted \\
 $\sin^{2}\theta_{23}$ & 0.50 &  Fitted \\
 $\Delta m^{2}_{32}$ & $2.4 \times 10^{-3}$ eV$^{2}$ & Fitted \\
 Mass hierarchy & Normal or Inverted & Fixed \\
 $\sin^{2}2\theta_{12}$ & 0.8704 & Fixed\\
 $\Delta m^{2}_{21}$ & $7.6 \times 10^{-5}$ eV$^{2}$ & Fixed\\
 \hline \hline
\end{tabular}
\end{center}
\label{fit_params}
\caption{Treatment of the oscillation parameters in the fitter.}
\end{table}

In order to have approximately the same number of expected events in neutrino and anti-neutrino modes, the run time ratio between $\nu$ and $\bar{\nu}$ was set to 1/3. However, it was shown that similar results are obtained for different ratios of neutrinos and anti-neutrinos. 

The idea of the Simple Fitter is to compare \textit{true} reconstructed neutrino energy spectra - the ones generated with \textit{true} oscillation parameters - with 
\textit{test} reconstructed energy spectra - the ones generated with \textit{test} oscillation parameters and to determine what value of free oscillation parameters 
would give the most similar \textit{test} energy spectrum with respect to the \textit{true}. 

Different values of the mixing parameters induce different oscillation probabilities which finally give different interactions in the far detector; thus this modifies the reconstructed neutrino energy spectra.

The fitter uses a binned maximum likelihood method to fit the reconstructed energy spectra as follows:
\begin{enumerate} 
 \item A set of reconstructed energy spectra are generated with ``true'' oscillation parameters, i.e. parameters that we choose as our reference.
 \item ``Trial'' oscillation parameters are generated and left fixed, where 1D fits use a single trial oscillation parameter and 2D fits use two trial oscillation 
parameters.  
 \item The remaining oscillation parameters are marginalized by fitting, which is done by minimizing the $\chi^{2}$ value comparing the ``true'' reconstructed energy 
spectrum to a reconstructed energy spectrum calculated with the trial and fitted oscillation parameters. 
 \item Systematics error nuisance parameters are marginalized by fitting which is done by minimizing $\chi^{2}$.
\end{enumerate}

Systematic errors are included by incorporating a realistic systematic error covariance matrix into the $\Delta \chi^{2}$ calculation.

The errors are estimated assuming the nominal oscillation parameters and are based on T2K. However, the expected improvements in the systematic errors related to the J-PARC and near detectors upgrades are taken into account.
\newpage
Systematic error sources are separated into three categories: 

\begin{enumerate}
 \item \textbf{Flux and cross section uncertainties constrained by the fit to near detector data}: We assume conservatively that this category of errors stays at the same level as T2K.
 \item \textbf{Cross section uncertainties not constrained by the fit to near detector data}: These uncertainties arise mainly from the fact that the cross section on water is not currently constrained by the near detector and the T2K replica target is not currently used. We assume that replica target data in NA61, that we use to reweight our beam MC, will be available and that cross section measurements will be made on water and these uncertainties will become negligible for Hyper-K.
 \item \textbf{Uncertainties on the far detector efficiency and reconstruction modeling}: Most of them are estimated using atmospheric neutrinos and the current precision is limited by statistics. Since Hyper-K has a bigger fiducial volume than Super-K, these errors are expected to decrease with more than an order of magnitude.
\end{enumerate}

\section{Results}

With a total exposure of 13\,MW $\times 10^{7}$ seconds integrated beam power, corresponding to $2.7 \times 10^{22}$ protons on target with 30\,GeV J-PARC beam, the CP phase, $\delta_{CP}$, can be determined better than 21 degrees for all possible values of $\delta_{CP}$ and CP violation can be established with a significance of more than 3$\sigma$ (5$\sigma$) for 78\% (62\%) of the $\delta_{CP}$ parameter space. 

The plots in Figure 3 compare these results (two tanks staging scenario) with a configuration where there would be only one tank or, on the contrary, three tanks. The tanks are all of the same dimensions.

%

\begin{figure}[h!]
\begin{center}
\begin{minipage}{0.48\linewidth}
\centering \includegraphics[scale=0.4]{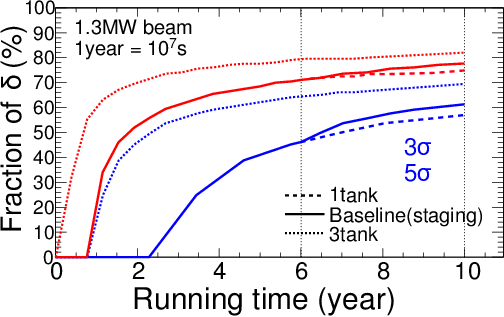}
\end{minipage}
\hfill
\begin{minipage}{0.48\linewidth}
\centering \includegraphics[scale=0.4]{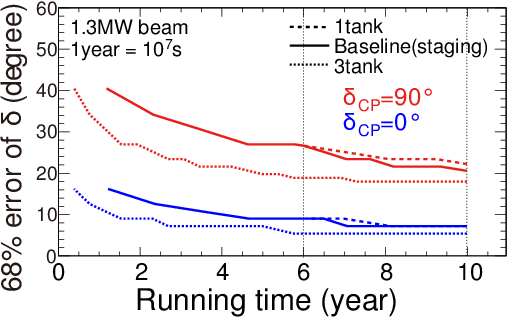}
\end{minipage}
\caption{Left: Fraction of $\delta_{CP}$ for which $\sin \delta_{CP} = 0$ (CP conserved case) can be excluded with more than 3$\sigma$ (red) and 5$\sigma$ (blue) significance as a function of integrated beam power. Right: Expected 1$\sigma$ uncertainty of $\delta_{CP}$ as a function of integrated beam power.}
\end{center}
\label{results}
\end{figure}

\section{Conclusion}
A design study was done to test Hyper-Kamiokande's sensitivity to the mixing parameters with different tank options; this resulted in selecting two tanks in a staging scenario. Sensitivity to the CP phase $\delta_{CP}$ is presented and the latest results are shown in Figure \ref{results}. 

This study is realized using the Simple Fitter, a binned maximum likelihood method with a systematic error covariance matrix. Systematic errors are treated in the same way as for T2K but the expected improvements related to the J-PARC and near detectors upgrades are included.

\end{document}